# Cognitive Twins for Supporting Decision-Makings of Internet of Things Systems

Jinzhi Lu[1], Xiaochen Zheng[1], Ali Gharaei[1], Kostas Kalaboukas[2], Dimitris Kiritsis[1]

[1] EPFL - École Polytechnique Fédérale de Lausanne, Lausanne, 1015, Switzerland
[2] SingularLogic SA, Athens, Greece
`dimitris.kiritsis@epfl.ch`

**Abstract.** Cognitive Twins (CT) are proposed as Digital Twins (DT) with augmented semantic capabilities for identifying the dynamics of virtual model evolution, promoting the understanding of interrelationships between virtual models and enhancing the decision-making based on DT. The CT ensures that assets of Internet of Things (IoT) systems are well-managed and concerns beyond technical stakeholders are addressed during IoT system development. In this paper, a Knowledge Graph (KG) centric framework is proposed to develop CT. Based on the framework, a future tool-chain is proposed to develop the CT for the initiatives of H2020 project FACTLOG. Based on the comparison between DT and CT, we infer the CT is a more comprehensive approach to support IoT-based systems development than DT.

## 1 Introduction

Internet of things (IoT) is a network of items embedded with sensors which are connected through the internet [1]. One IoT system consists of computing devices, physical plants and networks defined as a system-of-systems (SoS) [2]. During developing IoT systems, architectural dependencies across the entire SoS are challenged because of the massive compositions among them. During the lifecycle of IoT, virtual model assets for system, subsystems and components are needed to specify, detect and re- solve dependencies across domains, such as interface definition. Compositions from different domains and hierarchies of IoT system are evolving fast. Well managed and predictable evolution dynamics reduce the risks brought by new compositions, such as new characteristics and interoperability. Moreover, the architecture of IoT systems should be permitted with easy connectivity, control and communication among domain-specific applications. Thus, understanding the interrelationships between systems, subsystems and components is very important.

The motivation of our work is to overcome the challenges identified in the above paragraph and provide a new concept and framework to support IoT system development as follows. First, during IoT development, the virtual model asset should be managed in a systematic way during initial phases. An integrated information infrastructure with virtual models should enable to describe the interrelationships of IoT compositions to promote the understanding of their dependency and traceability. Second, the dynamics of model evolution need to be identified in order to predict the evolution of IoT system, subsystem and compositions. Third, topologies between virtual model assets enable to represent interrelationships of IoT compositions. Thus, the topologies are required to be managed.

Our contribution is to illustrate a new concept called Cognitive Twins (CT) and a Knowledge Graph (KG)-centric framework supporting CT development. We first define the concept of CT and digital twins (DT) to distinguish the differences between them. Then based on the concept of CT, a KG-centric framework is proposed to develop CT. Using KG, the topologies of virtual model assets are identified and managed. Moreover, a tool-chain concept is designed to support the framework for developing future CT. The results will be used in the H2020 projects FACTLOG[1] and QU4LITY[2].

The rest of the paper is organized as follows. We discuss the related work in Section 2 and introduce the definition of CT in Section 3. Moreover, the KG-centric framework is proposed in this section to create CT models. In Section 4, a future tool-chain concept is proposed for the related developments in the H2020 project FACTLOG. Finally, we discuss about CT in Section 5 and offer the conclusions with a summary in Section 6.

## 2    Related Work

The concept of DT was fostered by the rapid development of various existing technologies such as 3D modeling, system simulation, digital prototyping etc. [3]. In the whitepaper [4] published in 2014, Grieves defined the concept of DT and proposed a three-dimension model of DT based on the previous conception of "a virtual, digital equivalent to a physical product". According to Grieves, a DT model should at least consist of three main parts including: physical products in Real Space; virtual products in Virtual Space; and the connections of data and information that tie the virtual and real products together [4]. Since then, DT and relevant technologies have been evolving rapidly, which reflects that the virtual world and the physical world are becoming increasingly linked to each other and integrated as a whole [5]. Tao F. et.al. extended the existing three-dimension DT model by adding two more dimensions, DT data and services, and proposed a five-dimension model to promote the further applications of DT in more fields [6]. In a recent study, Qi Q. [5] et.al. reviewed the application fields, enabling technologies and tools for DT. Based on this study, it is concluded that universal design and development platforms and tools for DT are required to facilitate the integration of different technologies and tools which may have different formats, protocols and standards.

Data from different platforms and sources might be heterogeneous in syntax, schema, or semantics, which make data integration difficult. Semantic technologies provide solutions to achieve semantic interoperability in a heterogeneous system [7]. Semantic models enable to capture complex systems in an intuitive fashion, which can be summarized in standardized ontology languages, and come with a wide range of off-the- shelf systems to design, maintain, query, and navigate semantic models [8]. This characteristic makes semantic modelling a promising paradigm to address the challenges that DT development is facing currently. The authors of [8] employed semantic technologies to design a system that supports semantics-based DT. Many of existing researches use ontologies as the knowledge base, but the manual construction of ontologies is a very time-consuming task [9]. To overcome this limitation, more advanced techniques such as KGs are being used. According to [10][11] KGs acquire and integrate information into an ontology and utilize a reasoner to derive new knowledge and they can model information in the form of entities and relationships between them. KGs have been adopted in some studies to accelerate the implementation of DT. For example, in [12] the authors anticipated the paradigm of the next generation DT and KGs were considered as one of the main enabling technologies to link and retrieve all kinds of data, descriptive and simulation models etc. In [13], the authors analyzed the feasibility of backing DT with enterprise KGs based on the fact that DT could be strengthened by using semantic technologies to provide a formal representation of the DT domain. In [14] a graph-based query language was utilized to extract and infer knowledge from large scale production line data, to help generate DT models and therefore enhance manufacturing process management with reasoning capabilities.

---

[1] H2020 Project FACTLOG: http://factlog.eu/
[2] H2020 Project QU4LITY: https://qu4lity-project.eu/

Despite the importance of semantic technologies and KGs for the development of DT, there are still many gaps to be bridged, such as the lack of unified implementation architecture, integration of enabling technologies and tools etc. More research efforts are required for this topic.

## 3 Cognitive Twins

In this chapter, basic concepts of DT and CT are first introduced. Then the characteristics of IoT are introduced in order to formulate the problem of IoT systems. Then a KG- centric framework is proposed to construct CT for IoT systems.

### 3.1 Basic Concepts

In this chapter, concepts of DT and CT are introduced, as shown in Fig. 1, separately. Based on their respective concepts, the differences between them are summarized.

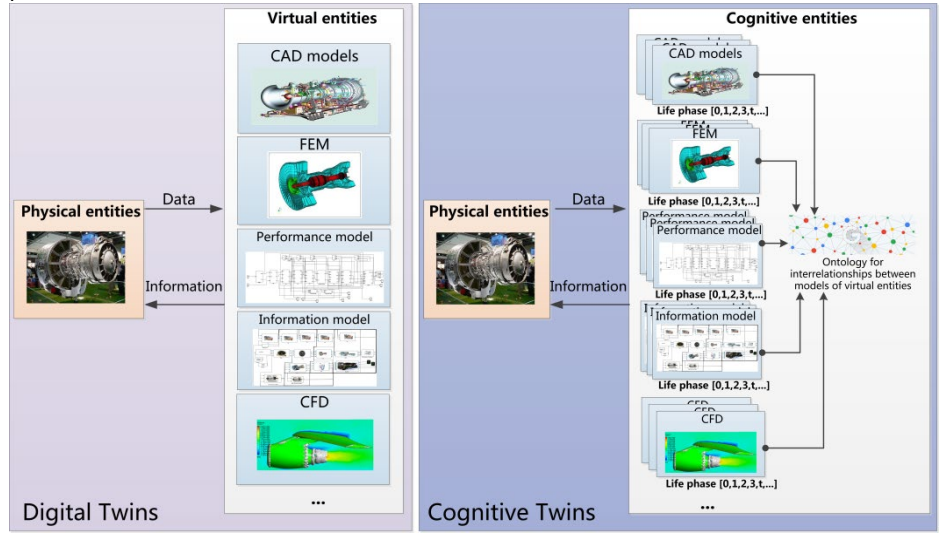

**Fig. 1**.Digital Twins vs Cognitive Twins

**Digital Twins (DT).** DT is a digital duplication of entities with real-time two-way communication enabled between the cyber and physical spaces [4]. It aims to support integration of IoT for connecting the physical and virtual spaces. As shown in Fig. 1, if the physical twin is defined as an areo-engine, the virtual entities of areo-engine include CAD models, FEM models etc. In this study, the concept of DT is formally defined as follows:

$$DT_{Sys}=PE\{Sys\} \cup VE\{\Sigma Model(Ms, Mp, Mt, Ml, Mt, Mm)\} \cup Comm\{\Sigma Data(EntitySt, EntityDe, Dtype, Datacontent)\} \quad (1)$$

where $DT_{Sys}$ refers to a DT of system *Sys*; *PE{Sys}* refers to the physical twin of *Sys*; *VE{ΣModel(Ms, Mp ,Mt, Ml, Mt, Mm)}* refers to a collection of models related to *Sys*. Each model includes several items:
- *Ms* (Model Structure): topology of models, inputs, outputs and parameters.
- *Mp* (Model purpose): the views of modeling, "why is the model needed?"
- *Mt* (Modeling theory): the mathematical foundation of modeling, e.g. differential-algebraic system of equations.
- *Ml* (Modeling language): any language expressing information or knowledge or systems in a structure that is defined by a consistent set of rules.
- Mt (Modeling tool): tools implementing models.
- *Mm* (Modeling method): a set of concepts to explain "how to develop models using a given language in one modeling tool to represent the formalisms?", e.g. fi- nite element modeling and structural equation modeling.

*Comm{ΣData(EntitySt, EntityDe, Dtype, Datacontnt)}* refers to data and information flows between physical entities and virtual entities. Each flow includes several items:
- *EntitySt* (Entities of Start): start of the data and information flow.
- *EntityDe* (Entities of Destination): destination of the data and information flow.
- *Dtype* (Type of data): type of data, such as real-time data and off-line data.

• *Datacontent* (Content of data): the data used in this data flow.

**Cognitive Twins (CT).** DTs are expected to support the industrial area of design, production, prognostics, and health management, etc. [15]. Each DT has different models which are difficult to manage, because the model versions are updated across the lifecycle. Moreover, the virtual models in DT are across domains which are difficult to identify their interrelationships. The CT is proposed to solve this problem as shown in Fig. 1. One timestamp for each lifecycle spot is added to each virtual model. Moreover, topologies of models are required to be described.

$$CT_{Sys}=PE\{Sys\} \cup VE\{\Sigma Model_t(Ms_t, Mp_t, Mt_t, Ml_t, Mt_t, Mm_t), Ontopology(entities, relationships)\} \cup Comm\{\Sigma Data(EntitySt, EntityDe, Dtype, Datacontent)\}, \quad t=1,2,3,\ldots, timespots\ in\ lifecycle \qquad (2)$$

Where $CT_{Sys}$ refers to a CT of system *Sys*; $PE\{Sys\}$ refers to the physical twin of *Sys*; $VE\{\Sigma Model_t(Ms_t, Mp_t, Mt_t, Ml_t, Mt_t, Mm_t), Ontopology(entities, relationships)\}$ refers to a collection of models related to *Sys*. Different from DTs, each model in the CT is added with a timestamp in the lifecycle. Except for the items in DTs, *Ontopology (entities, relationships)* refers to ontology to represent the *topology between Models*.
 • The *entities* refer to all the information related to models, such as compositions.
 • The *relationships* refer to all the interrelationships of entities.

### 3.2   Problem Formulation

Based on the basic concept of proposed CT, a KG-centric framework is proposed for supporting our EU Projects FACTLOG and QU4LITY and Swiss InnoSwiss IMPULSE project on DTs. These three projects are mainly focusing on IoT systems using DT. Based on the initiative definition [16], several technological and social aspects related to IoT are investigated to identify the industrial concerns for developing the framework in the next section.

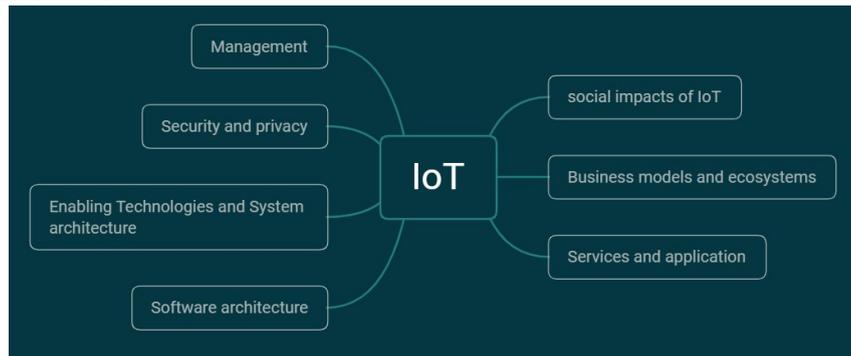

**Fig. 2.** IoT concerns

As shown in Fig. 2, seven aspects are considered during the entire lifecycle of IoT. The details are introduced as follows:
 • Social impacts of IoT, such as impacts and acceptance of users.
 • Business models and ecosystems, a new business model for IoT systems.
 • Services and application, including domain specific services.

- Software architecture, such as operational systems, middleware.
- Enabling technologies and systems architecture, sensors, energy management
- Security and privacy, such as management of personal data.
- Management, such as autonomics and self-organization of large IoT systems.

### 3.3 A KG-centric Framework for Cognitive Twins

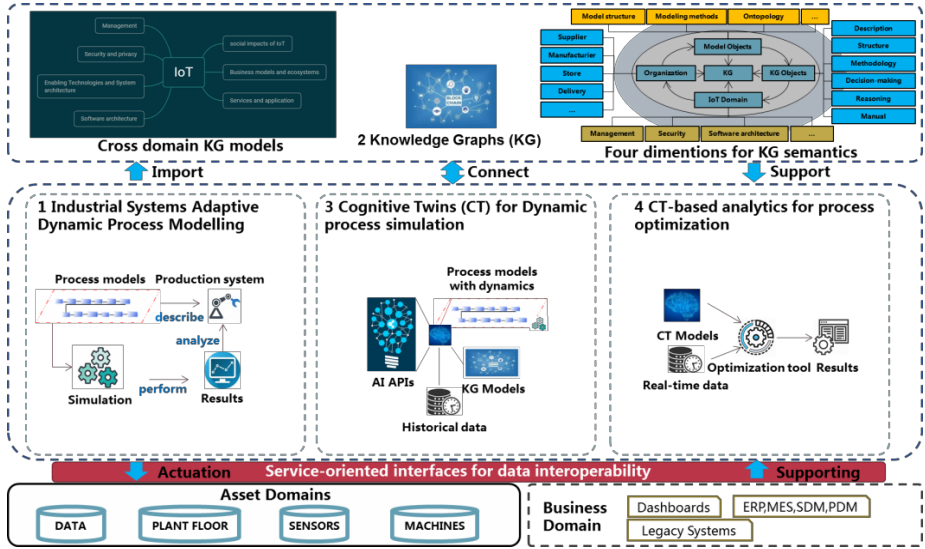

**Fig. 3. A KG-centric Framework for CT**

In order to construct CT for IoT systems, a KG-centric framework is proposed as shown in Fig. 3. It requires inputs from business domains and performs outputs to asset domains. The five main patterns of the KG-centric framework are shown in details as follows:

**Process Modeling and Simulation.** IoT systems consist of computational compositions, sensors, networks and plants which are considered as hybrid systems including continuous systems and discrete systems [17]. DT is an integrated system consisting of mathematical models and data, which is closed to a real-time synchronization be- tween real physical systems and their own virtual entities [18]. Such process can be represented as entire workflows where the computing composition and other plant nodes are linked together. In this pattern, a process modeling and simulation approach is used to formalize these workflows and to simulate the hybrid system behaviors.

**Ontology-based Knowledge Graph.** KG models are at core to represent the topological interrelationships between physical entities and cognitive entities. Before developing KG models, ontologies for KG models are first designed in order to develop the semantics and syntax. Based on the basic concepts of CT and problem formulations in Section 3.3, the ontology includes:

- **IoT domains.** This part focuses on the contents related to IoT domains including physical entities and communications. Seven aspects in Section 3.2 are considered when defining the ontology.

- **Model objects.** This part mainly focuses on the contents related to CT, such as model structure and *Ontopology* (topology between models).
- **Organizations.** This part mainly focuses on the organizations related to IoT,
- such as suppliers and stores.
- **KG objects.** This part mainly focuses on the knowledge graphs including description, structure, methodology, decision-making, reasoning and manuals.

**Cognitive Twins for Dynamic Process Simulation.** Artificial Intelligence (AI) APIs, KG models, historical data and process models with dynamics are integrated to generate CT models. CT models aim to support decision-makings for dynamic processes of physical entities.

**CT-based Analytics for Process Optimization.** Based on the CT models and real-time data, a tool is used to support process optimization. The optimization results are performed to make decisions for manipulating the physical entities.

**Service-oriented Interfaces for Data Interoperability.** A service-oriented approach is proposed to develop interfaces for heterogeneous data. All the assets and business domain data are transformed to unified formats through the developed interfaces. Such unified data are used to support other patterns in the framework.

## 4   A Future Tool-chain for Developing Cognitive Twins

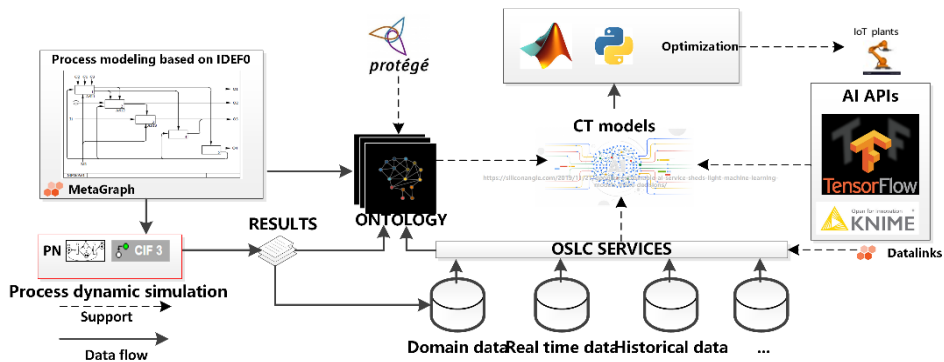

**Fig. 4.** Overview of the tool-chain

Based on the proposed KG-centric framework, a tool-chain is developed for developing the CT models in which several tools are adopted as compositions of the tool- chain, as shown in Fig. 4. The detailed tools are introduced in Table 1. *MetaGraph* is a DSM tool to develop process models based on *IDEF0* [19]. Moreover, *MetaGraph* generates CIF models for process dynamic simulation for process models. The do- main data including process dynamic simulation results, real-time data and historical data of IoT systems are represented as Open Services for Lifecycle Collaboration (OSLC) services through datalinks (a tool for developing OSLC adapters). The OSLC services are RESTful services for linking data through defined URIs. Moreover, we make use of Protégé to formalize process models, process dynamic simulation results and OSLC services from domain data. AI APIs including TensorFlow and KNIME are used to generate CT models based on ontology and OSLC services. The CT models are used for supporting optimize the IoT plants which optimization algorithms are developed using Matlab and Python.

Table 1. The initial tool-chain for developing CT models

| Tools | Descriptions |
|---|---|
| MetaGraph[3] & MetaEdit+ [20] | Process modelling |
| CIF simulator [21] & PN simulator [22] | Process dynamic simulation |
| Protégé [23] | Ontology modeling |
| KNIME [24] & Tensorflow [25] | Develop CT models |
| Matlab [26] & Python [27] | Design optimization toolset for the dynamic processes |
| Datalinks[4] | Developing OSLC services for domain data [28] |

## 5   Discussion

Currently, DTs are proposed to support the entire lifecycle of IoT. Physical entities, virtual entities, data, service and connections between them are always concerned by industries, such as NASA [15]. From the literature review, traditional DTs at core focus on connections between the physical entities and virtual entities. The main difference between DT and the proposed CT are replacing the virtual entities by CTs. The CTs add timestamp for each model and provide topologies between all the models. Thus, the cognitive models are dynamically evolved rather than being static according to the physical entities. Several use cases are defined when CT is used:

Table 2. The initial tool-chain for developing CT models

| Use case | Description |
|---|---|
| Lifecycle dynamics | The added timestamp for each model is used to analyze the dynamics of virtual model evolutions. |
| Decision-makings | The lifecycle dynamics provide clues for decision-makings for the system evolution. |
| Data analysis across domains | The topology of virtual entities provides a unified description of across domain data which is the basis for data analysis at entire system level. |

Taking an example of aero−engines, DTs are used for constructing the prognostic health management system, which the physical engine is connected with the digital models in order to realize real-time aero−engine monitoring and fault detection. However, the lifecycle of aero−engine is very long leading to that there are various versions of models used before the aeroengine is finalized. Moreover, the aeroengine consists of different compositions which are used for different scenarios of production, operation and maintenances. The topologies between different virtual models with different versions, domains and hierarchies identify the lifecycle dynamics and domain interrelationships of each model which provide clues about dynamics of system lifecycle and a decision-making solution based on system-level data. Thus, several advantages are summarized:

- The time stamps for each model of CTs promote the dynamics of the virtual model evolution. Based on this dynamics, decision-making based on the CTs enable to predict not only the behaviors of physical entities, but also the model updates of the virtual entities (concepts in DTs).
- Ontology for representing interrelationships between models also provides more clues for analyzing the behaviors of physical entities.

---

[3] A Domain-Specific Modeling tool of Z.K. Fengchao http://www.zkhoneycomb.com/
[4] A tool for developing OSLC services [27] of Z.K. Fengchao http://www.zkhoneycomb.com/

This paper focuses on IoT system development, operation and maintenance. The IoT system developers expect to have a good dependency from requirement, function, behaviors and architecture when developing IoT systems. Moreover, the lifcecycle of IoT systems is shorter than traditional equipment, such as areo-engine. The components are renewed quickly which means the entire IoT systems evolve fast. Further- more, IoT requires flexible and standardized interfaces during they are developed because of such fast evolutions.

Based on the summarized advantages of CT, ontology promotes the understanding of dependencies between models, such as requirement models. In order to support fast evolution of IoT systems, dynamics of virtual models are useful to analyze the system changes and to identify the requirements for new system components. The flexible and standardized interfaces also require a good understanding of interrelationships between physical components or between models. Totally, CT has the better capabilities to support IoT system development compared with DTs.

## 6  Conclusion

This paper presents a conceptual definition of CTs supporting IoT system development and maintenance. Based on the definition, a knowledge graph based framework is proposed to develop CT models. Based on the framework, a future tool-chain concept is used to support an initiative solution for the H2020 project FACTLOG.

## 7  Acknowledgement

The work presented in this paper was supported by the EU H2020 project (869951) FACTLOG-Energy-aware Factory Analytics for Process Industries and EU H2020 project (825030) QU4LITY Digital Reality in Zero Defect Manufacturing and the InnoSwiss IMPULSE project on Digital Twins.